\documentclass[a4paper,12pt,leqno]{article}

\usepackage{graphicx}
\usepackage{subeqn}
\usepackage{amssymb}
\usepackage{lineno}
\usepackage{bm}
\usepackage{array}
\usepackage{color}
\usepackage{subfigure}

\date{ }

\begin{document}

\title{Statistics of non-linear stochastic dynamical systems under L\'evy noises by a convolution quadrature approach}

\author{Giulio Cottone $^{1,2}$ \thanks{E-mail: giuliocottone@unipa.it; giulio.cottone@tum.de }\\
\small Universit\'a degli Studi di Palermo, Italy\\
\small $^2$ Engineering Risk Analysis Group, Technische Universit\"at M\"unchen, Germany
}


\maketitle


\begin{abstract}
This paper describes a novel numerical approach to find the statistics of the non-stationary response of scalar non-linear systems excited by L\'evy white noises. The proposed numerical procedure relies on the introduction of an integral transform of Wiener-Hopf type into the equation governing the characteristic function. Once this equation is rewritten as partial integro-differential equation, it is then solved by applying the method of convolution quadrature originally proposed by Lubich, here extended to deal with this particular integral transform.
The proposed approach is relevant for two reasons: 1) Statistics of systems with several different drift terms can be handled in an efficient way, independently from the kind of white noise; 2) The particular form of Wiener-Hopf integral transform and its numerical evaluation, both introduced in this study, are generalizations of fractional integro-differential operators of potential type and Gr\"unwald-Letnikov fractional derivatives, respectively.
\end{abstract}

\section{Introduction}
Non-linear stochastic differential equations (SDEs) model many real phenomena.
Since the early works of Brown and Langevin, SDEs driven by Gaussian white noise processes have been studied and for long time they have represented the standard approach in stochastic modeling. 
Only in the last decades, L\'evy white noise excited systems have attracted the interest of many authors in finance, in physics and in engineering.
The reason of such an interest is that many physical phenomena exhibit non-Gaussian behavior which cannot be neglected: skewed probability distributions, inverse power-law decay of the density and divergent moments are typical characteristics which cannot be modeled by Gaussian distribution.

The generalized form of the central limit theorem given by Paul L\'evy provides the necessary mathematical support to justify the choice of non-Gaussian excitation, stating the conditions under which the sum of independent random variables converges to the so-called L\'evy $\alpha$-stable distribution.
This is a class of distributions defined in terms of the stability index $0<\alpha \le 2$, being $\alpha=2$ the Gaussian distribution.

If on the one hand non-Gaussian assumption is useful to enriching models which become more and more realistic, on the other hand this increases the mathematical complexity of the governing equations. 
For example, the Fokker-Planck equation, regarding the time evolution of the probability density function, is a partial differential, an integro-differential or a fractional differential equation in the cases of Gaussian, Poisson and $\alpha$-stable white noise external excitation, respectively. 
Very few analytical solutions are available and numerical methods to find the statistics of the solution of the non-linear SDE are frequently the only valid alternative.

In this context, this paper introduces a relevant numerical method for handling with this problem in case of externally driven scalar SDEs. 
It will be shown that the proposed method applies to every kind of L\'evy white noise external excitation and that it is numerically advantageous. 
To frame this method, it is useful to briefly recall some approach given in literature, considering separately the Gaussian, the Poisson and the $\alpha$-stable L\'evy case.

SDEs excited by Gaussian white noise have been studied for a long time and, for some particular form of the drift and diffusion coefficients, stationary density \cite{lin67, sobc91} are available by solving the associated Fokker-Planck-Kolmogorov (FPK) equation. 
Non-stationary solution can also be, at least theoretically, found \cite{risk96} if the eigenfunctions of the non-linear FPK are available or can be computed in numerical way. Among the proposed strategies to approach the case of Gaussian excitation we mention: 
1) methods to solve the FPK equation \cite{masu05, spen85, spen93, berg93, wedi99} by numerical approaches, in particular by Finite Element approach; 
2) path integral method \cite{naes93}; 
3) methods to transform the original non-linear systems into an equivalent systems, which allow the approximation of high dimensional systems, such as the stochastic linearization \cite{robe99} and the stochastic averaging \cite{khas66}.  
While methods 1) and 2) are confined to dynamical systems with relatively small number of state variables and are still impracticable \cite{masu05a} in high dimensions, methods in 3) may not be accurate in case of strong non-linear drift.

For Poisson white noise excitation, generalizations of the Fokker-Planck equation for the evolution of the response density, indicated as Kolmogorov-Feller (KF) equation, are given in \cite{dipa93, dipa93b, gihm72} and they are based on a generalized form of the It\^o's calculus; another approach starting from the theory on filtered Poisson processes \cite{grig96b} leads to the equations governing the characteristic function in case of polynomial drift and external Poisson white noise, that are the spectral counterpart of the KF equation. 
Exact stationary solutions for few cases have been proposed in \cite{prop03}. Numerical procedures to find statistics \cite{bara07}, equivalent linearization \cite{grig95}, path integral solution \cite{dipa08, pirr11} and extensions of the Bogolyubov's theorem (stochastic averaging) \cite{kolo91} are also available.

The study of L\'evy white noise processes is characterized by mathematical complexity mainly due to the heavy tails of the $\alpha$-stable distribution, which imply that moments of order $p>\alpha$ are divergent. 
This causes the inapplicability of many standard procedures of stochastic calculus, based on moments.
It\^o's calculus has been extended to deal with L\'evy white noise processes by many authors and readers are referred to \cite{prot04} for a rigorous mathematical treatment and to \cite{grig02} for a complete presentation of the topic with applications in sight.
Although this topic attracted interest only in the last decades, a considerable amount of research has been conducted: analytical stationary solution for particular polynomial potential are given in \cite{chec02, chec04, dybi10, gonc02}; equivalent linearization technique based on the minimization of a functional of the characteristic function is presented in \cite{grig00}; methods based on the continuous time random walk leading to the generalization of the Fokker-Planck equation in terms of fractional Riesz derivatives can be found in the topic reviews \cite{metz00, metz04}.

To the author's knowledge none of the above cited method is applicable, without substantial modification, to the treatment of SDEs excited by any external L\'evy noise. 
This paper fills this gap showing a numerical method to find the statistics of the non-stationary response of non-linear systems driven either by Gaussian, Poisson or $\alpha$-stable white noise processes.

The solution strategy is based on a proper numerical treatment of the equations ruling the characteristic function that are indicated as Spectral Fokker-Planck equations (SFPEs). 
Main advantage of these SFPEs is that their mathematical form remains unchanged depending on the excitation. 
Although dealing with the characteristic function rather than with the probability density is a strategy already used by other authors, see \cite{chec04}, an unique numerical treatment for any L\'evy white noise has never been proposed, to the best of the author's knowledge. 
Moreover, it is important to stress that this method is not restricted to drift coefficient of polynomial type, or to power-law type as in \cite{chec04}, but conversely is extremely advantageous and straightforwardly applicable to any non-linear form of the drift, as the numerical section shows.
To this aim, we will give firstly a computationally suitable form to the governing equations in order to highlight the integro-differential structure of the equations. Then, once the integro-differential nature of the SFPEs is made explicit, a Wiener-Hopf integral transform will be introduced and the convolution quadrature method, originally proposed by Lubich \cite{lubi88a, lubi88b, lubi04}, is reformulated in order to handle such an operator. 
In this way, the characteristic function of the response of non-linear stochastic differential equation is found solving a simple linear system of ordinary differential equations.  

It is worth to mention that one of the strength of the proposed approach relies on the use of convolution quadrature method, that here is properly modified to suit our problem in a straightforward way. 
Indeed, as pointed out in \cite{scha06} convolution quadrature has excellent stability properties compared to other discretization method and is a valid alternative both from the accuracy and the efficiency point of view.

\section{Problem statement}

Consider the nonlinear system excited by an external white noise process, in the form 
\begin{equation}
\left\{\begin{array}{l} {\dot{X}\left(t\right)=f\left(X\left(t\right),t\right)+W_{L} \left(t\right)} \\ {X\left(0\right)=X_{0} } \end{array}\right.  
\label{eq_1}
\end{equation}
where $f\left(X\left(t\right),t\right)$ is non-linear deterministic function of the response process $X(t)$ and $X_{0} $ is an initial condition, which might be either deterministic or random, with assigned distribution. In the framework of the generalized theory of random processes a stationary white noise process $W_{L} (t)$ is defined as the formal derivative of a process with stationary orthogonal increments $L(t)$ and starting from zero a.s., i.e. $L(0)=0$, called L\'evy process. 
Formally we can write
\[
W_{L} \left(t\right)\mathop{=}\limits^{def} \frac{{\rm d}L\left(t\right)}{{\rm d}t} 
\] 
that, introduced into (\ref{eq_1}) gives the It\^o's form 
\begin{equation}
dX\left( t \right) = f\left( {X\left( t \right),t} \right)dt + dL\left( t \right)
\label{eq_1b}
\end{equation}

In what follows we restrict our attention to the following three interesting cases of external excitation:

\begin{enumerate}
\item  Gaussian or normal white noise process $W_{G} (t)$ defined as the formal derivative of the Brownian motion $B(t)$ whose increments $dB(t)=B(t+dt)-B(t)$ have zero mean Gaussian distribution and with constant power spectral density equal to $q/2\pi $, where $q$ indicates the noise strength, assumed constant in time. 

\item  Poisson white noise process $W_{P} (t)$ defined as the formal derivative of the Compound motion $C(t)$; $W_{P} (t)$ is defined as sum of Dirac's delta impulses $\delta (t-T_{k} )$, having random distributed amplitudes $Y_{k} $, occurring at Poisson distributed random times $T_{k}$ independent from $Y_{k} $, explicitly 
\[
W_{P} \left(t\right)\mathop{=}\limits^{def} \sum _{k=1}^{N\left(t\right)} Y_{k} {\kern 1pt} \delta \left(t-T_{k} \right)
\]  
where $N(t)$ indicates a counting process with mean arrival rate $\lambda$. 
In the following, it is assumed that the distribution of the random variable $Y$ is known along with its characteristic function $\phi _{Y}(\theta )$. 
No further assumptions are done regarding the existence of moments of $Y$.

\item  $\alpha $-stable L\'evy white noise process $W_{\alpha }(t)$ defined as the formal derivative of the $\alpha $-stable L\'evy motion $L_{\alpha } (t)$, see \cite{grig02}.
\end{enumerate}

By characterizing the driving noise, it is possible to derive the equation governing the evolution of the characteristic function, that we indicate as Spectral-Fokker-Planck equations (SFPEs) (see \cite{cott10c}). 
For stochastic differential equations excited by external Gaussian, Poisson and $\alpha $-stable L\'evy white noise processes, SFPEs are written, omitting arguments, as
\begin{subequations}\label{eq_2}
\begin{eqnarray}
\dot{\phi }={\rm{i}}\theta E\left[f\left(X,t\right)e^{{\rm{i}}\theta X} \right]-\frac{\theta ^{2} }{2} \phi  
\label{eq_2a}
\end{eqnarray}
\begin{eqnarray}
\dot{\phi }={\rm{i}}\theta E\left[f\left(X,t\right)e^{{\rm{i}}\theta X} \right]-\lambda {\kern 1pt} {\kern 1pt} \phi \left(1-\phi _{Y} \right)
\label{eq_2b}
\end{eqnarray}
\begin{eqnarray}
\dot{\phi }={\rm{i}}\theta E\left[f\left(X,t\right)e^{{\rm{i}}\theta X} \right]-\left|\theta \right|^{\alpha } \phi 
\label{eq_2c}
\end{eqnarray}
\end{subequations}
where ${\rm{i}} = \sqrt{-1}$ is the imaginary unit, $E\left[\cdot \right]$ denotes expectation and $\phi =E\left[e^{{\rm{i}}\theta X\left(t\right)} \right]$ is the characteristic function of the response $X(t)$. 
Readers are referred to \cite{grig04, samo07} and to \cite{dipa07} for different prospectives to derive the characteristic function equations of the state of dynamical systems driven by L\'evy white noises.

From the above SFPEs it is noticed that the non-linear drift term influences the time evolution of the characteristic function only in the term $E\left[f\left(X,t\right)e^{{\rm{i}} \theta X\left(t\right)} \right]$.

The case in which $f\left(X,t\right)$ is polynomial of the state variable $X(t)$ is handled easily.
In such a case Eqs.(\ref{eq_2}) become partial differential equations by the property 
\begin{equation} 
\label{eq_3} 
\frac{\partial ^{k} \phi \left(\theta ,t\right)}{\partial \theta ^{k} } = E\left[\left({\rm{i}} X\right)^{k} e^{{\rm{i}} \theta X\left(t\right)} \right] 
\end{equation} 
which follows from the definition of $\phi \left(\theta, t\right)$ and standard numerical schemes for the solution of partial differential equations can be used.

On the contrary, if $f(X,t)$ is not a polynomial in $X(t)$, the term $E\left[f(X,t)e^{{\rm{i}} \theta X\left(t\right)} \right]$ is not directly related to the function $\phi (\theta ,t)$.

Aim of this paper is to provide a general numerical framework to solve the It\^o's stochastic differential equation (\ref{eq_1b}) in which the non-linear drift term is not of polynomial form. 
The strategy used is based on the solution of eqs.(\ref{eq_2}).
First, the expectation $E\left[f(X,t)e^{{\rm{i}}\theta X\left(t\right)} \right]$ is rewritten as a Wiener-Hopf convolution integral operator applied to the characteristic function. 
Then, the Lubich's method for the quadrature of convolution integral will be adapted to find a form that is numerically convenient.
Finally, the general problem of numerical evaluation of such an integral by convolution quadrature will be addressed and a procedure for the numerical evaluation of Eqs.(\ref{eq_2}) will be proposed and tested by some examples.

\section{Convolution quadrature for the integro-differential equation governing the characteristic function}
\noindent Indicate with $X(t)$ the response to the stochastic differential equation in (\ref{eq_1}) and let $p(x)$ and ${\phi(\theta)}$ denote the density and the characteristic function of $X(t)$, respectively, where the dependence on time is for readability's sake not explicitly indicated.

From the definition of the expectation operator and recalling that the characteristic function is the Fourier transform of the density $p(x)$, the expectation $E\left[ {f\left( X \right)e^{i\theta X} }\right]$ can be written as
\begin{eqnarray}
		E\left[ {f\left( X \right) e ^{i\theta X} } 
		\right]
		 = \int_{ - \infty }^\infty  {p\left( x \right)f\left( x 
		 \right)e^{i\theta
		  x} dx}  =  \nonumber \\ 
		    = \frac{1}{{2\pi }}\int_{ - \infty }^\infty  {\int_{ - \infty 
		    }^\infty 
		     {\phi \left( u \right)e^{ - iux} du\,} f\left( x 
		     \right)e^{i\theta x}
		      dx}  =  \nonumber \\ 
		        = \frac{1}{{2\pi }}\int_{ - \infty }^\infty  {\phi \left( u
		         \right)\int_{ - \infty }^\infty  {f\left( x \right)e^{ - i\left( 
		         {u - \theta } \right)x} dx} \,du}   
		\label{equ4}
\end{eqnarray}

The inner integral is the Fourier inverse transform of $f(x)$, which can be indicated by $\hat{f}(u)$, i.e.
\begin{equation}
		\hat{f}(u) = \frac{1}{{2\pi }}\int_{-\infty }^\infty 
		 {f\left( x \right)e^{ - i u x} dx} \nonumber 
		 \label{equ5}
\end{equation}

Introducing the latter in eq.(\ref{equ4}), the relation
\begin{equation}
E\left[ {f\left( X \right)e ^{i\theta X} } 
		\right]= \int_{ - \infty }^\infty  {\phi \left( u \right)\hat{f}\left( {u -
		            \theta } \right)du}
\label{eq5a}
\end{equation}
in terms of the characteristic function $\phi(\theta)$ is obtained. This is an integral transform that can be referred to as Wiener-Hopf type, in analogy to the Wiener-Hopf integral equations. 
For what follows, it is more convenient to dub the previous convolution integral as $\left( {T_f\phi } \right)\left( \theta  \right) $ that is an integral transform applied to the characteristic function $\phi(\theta)$ and depending on the function $f(x)$. 

The advantage of this representation relies on the easy and efficient numerical evaluation of the transformation operator $(T_f\phi)(\theta)$ by the convolution quadrature shown in the following. This is adapted from the original work of Lubich \cite{lubi88a, lubi88b, lubi04}, which readers are referred to.

Let us indicate by
\begin{eqnarray}
 F_ -  \left( s \right) = \int_0^\infty  {\hat{f}\left( { - t} \right)e^{ - s t} dt} \nonumber 
\end{eqnarray}
and 
\begin{eqnarray}
 F_ +  \left( s \right) = \int_0^\infty  {\hat{f}\left( t \right)e^{ - s t} dt}  \nonumber 
\end{eqnarray}
the Laplace transform of the function $\hat{f}(t)$ for negative and positive values of the variable $t$, respectively. Moreover, let us recall the inverse Laplace transform relations 
\begin{subequations}\label{eq12}
\begin{eqnarray}
\hat{f}\left( t \right) = \frac{1}{{2\pi i}}\int_\Gamma ^{} {F_+\left( s \right)} \,e^{s t} ds
\label{eq12a}
\end{eqnarray}
\begin{eqnarray}
\hat{f}\left( -t \right) = \frac{1}{{2\pi i}}\int_\Gamma ^{} {F_-\left( s \right)} \,e^{s t} ds
\label{eq12b}
\end{eqnarray}
\end{subequations}

The latter can be introduced into the definition of the operator $(T_f\phi)(\theta)$, and after some algebraic manipulation one obtains 
\begin{eqnarray}
\left( {T_f\phi } \right)\left( \theta  \right) = \int_0^\infty  {\hat{f}\left( { - t} \right)\phi \left( {\theta  - t} \right)dt}  + \int_0^\infty  {\hat{f}\left( t \right)\phi \left( {\theta  + t} \right)dt} 
\label{eq13}
\end{eqnarray}

Introducing (\ref{eq12}) in the latter and rearranging, produces
\begin{eqnarray}
 \left( {T_f \phi } \right)\left( \theta  \right) = \frac{1}{{2\pi i}}\int_\Gamma  {F_ -  \left( s \right)\int\limits_{ - \infty }^\theta  {\phi \left( u \right)e^{s\left( {\theta  - u} \right)} duds + } }  \\ \nonumber
 \,\,\,\,\,\,\,\,\,\,\,\,\,\,\,\,\,\,\,\,\, + \frac{1}{{2\pi i}}\int_\Gamma  {F_ +  \left( s \right)\int\limits_\theta ^\infty  {\phi \left( u \right)e^{s\left( {u - \theta } \right)} duds} } 
\label{equ14}
\end{eqnarray}

The inner integral in the first term can be thought as the formal solution to the differential equation
\begin{equation}
\left\{ {\begin{array}{l}
   {y'\left( \theta  \right) =  - s\,y\left( \theta  \right) - \phi \left( \theta  \right)}  \\
   {y\left( -\infty  \right) = 0}  \\
\end{array}} \right.
\label{equ15}
\end{equation}
whose solution will be denoted by $y(\theta, s)$ to remark the dependency on the variable $s$ of the Laplace transform. In similar fashion, the second term can be expressed as the formal solution to the differential equation
\begin{equation}
\left\{ {\begin{array}{l}
   {z'\left( \theta  \right) =  - s\,z\left( \theta  \right) - \phi \left( \theta  \right)}  \\
   {z\left( \infty  \right) = 0}  \\
\end{array}} \right.
\label{equ16}
\end{equation}
whose solution is $z(\theta,s)$. 
Eq.(\ref{equ14}) can be rewritten as 
\begin{equation}
\left( {T_f\phi } \right)\left( \theta  \right) = \frac{1}{{2\pi i}}\int_\Gamma ^{} {F_ -  \left( s \right)y\left( {\theta ,s} \right)ds}  + \frac{1}{{2\pi i}}\int_\Gamma ^{} {F_ +  \left( s \right)z\left( {\theta ,s} \right)ds} 
\label{equ17}
\end{equation}

It is worth to note that the term $E\left[ {f\left( X \right)e^{i\theta X} } \right]$, taking into account of the non-linearity of the system, is thus expressed as an inverse Laplace transform of a functions depending of the unknown characteristic function $\phi(\theta)$. 
The main advantage of this representation relies on the suitable numerical solution that can be performed following the work of Lubich \cite{lubi88a, lubi88b} on quadrature of convolution integrals. 
The main point, outlined in the following, is that the solutions to the differential equations in eq.(\ref{equ15}) and eq.(\ref{equ16}) are expressed in terms of sums by applying difference schemes.
The following way to derive a series approximation slightly differs from the original method of Lubich that uses formal power series and operational calculus.
Main motivation is that eqs.(\ref{equ15}) and (\ref{equ16}) have an initial condition at infinity that can be taken into account by using recursive relations as shown below.
 
Eqs.(\ref{equ15}) and (\ref{equ16}) will be approximated by means of a simple Eulerian scheme for the first order derivative. 
Higher approximation might be pursued by higher order derivative schemes, but to keep a simpler notation, we will confine our attention to the simple Eulerian one.

Let us indicate $\theta _n  = nh$, $y\left( {\theta _n } \right) = y_n $, $z\left( {\theta _n } \right) = z_n $, $\phi \left( {\theta _n } \right) = \phi _n $, where $h \in \mathbb{R}$ is a  grid step and $n \in \mathbb{N}$.

The approximation of eq.(\ref{equ15}) can be pursued by a forward Eulerian scheme that reads
\begin{equation}
y_n  - y_{n - 1}  = s h y_n  + \phi _n h
\label{equ18}
\end{equation}

To find a solution to the latter equation, define the infinite dimensional vector 
${\bf{Y}} = \left[ {\begin{array}{*{20}c}
   {y_{n} } & {y_{n-1} } & {...}  \\
\end{array}} \right]^T $, collecting the solution at sampling steps, the infinite dimensional vector ${\bf{\Phi }} = \left[ {\begin{array}{*{20}c}
   {\phi _{n} } & {\phi _{n-1} } & {...}  \\
\end{array}} \right]^T$
and the formal matrix
\begin{eqnarray}
{\bf{A}} = \left[ {\begin{array}{*{20}c}
   0 & 1 & 0 & {}  \\
   0 & 0 & 1 &  \cdots   \\
   0 & 0 & 0 & {}  \\
   {} &  \vdots  & {} &  \ddots   \\
\end{array}} \right]
 \nonumber
\label{matrA}
\end{eqnarray}
such that the infinite set of equations represented by eq.(\ref{equ18}) can be written in the form
\begin{equation}
{\bf{Y}} = \left( {1 - s h} \right)^{ - 1} {\bf{A Y}} + {\bf{\Phi }}h\left( {1 - s h} \right)^{ - 1} 
\label{equ19}
\end{equation}

Formal solution to the latter is provided in the form
\begin{equation}
{\bf{Y}} = \left( {{\bf{I}} - \left( {1 - sh} \right)^{ - 1} {\bf{{A}}}} \right)^{ - 1} {\bf{\Phi }}h\left( {1 - sh} \right)^{ - 1} 
\label{equ20}
\end{equation}
being
\begin{eqnarray}
{\bf{I}} = \left[ {\begin{array}{*{20}c}
   1 & 0 &  \cdots   \\
   0 & 1 & {}  \\
    \vdots  & {} &  \ddots   \\
\end{array}} \right] \nonumber
\label{matrI}
\end{eqnarray}
an infinite dimensional identity matrix. Because of the particular structure of the matrix $\left( {{\bf{I}} - \left( {1 - s h} \right)^{ - 1} {\bf{{A}}}} \right)$, it is not hard to prove by induction that its inverse has the form
\begin{eqnarray}
  \left( {{\bf{I}} - \left( {1 - s h} \right)^{ - 1} {\bf{{A}}}} \right)^{ - 1}  = \left[ {\begin{array}{*{20}c}
   1 & {\left( {1 - s h} \right)^{ - 1} } & {\left( {1 - s h} \right)^{ - 2} } & {\left( {1 - s h} \right)^{ - 3} } &  \cdots   \\
   0 & 1 & {\left( {1 - s h} \right)^{ - 1} } & {\left( {1 - s h} \right)^{ - 2} } &  \cdots   \\
   0 & 0 & 1 & {\left( {1 - s h} \right)^{ - 1} } &  \cdots   \\
    \vdots  & {} & {} & {} &  \ddots   \\
\end{array}} \right]
\label{equ21}
\end{eqnarray}

Concluding the reasoning, the general term of the differential equation $y_n$ follows from eqs.(\ref{equ20}) and (\ref{equ21}) and after some algebraic manipulation it can be written as 
\begin{equation}
y_n  = \sum\limits_{k = 0}^\infty (-1)^{k+1} {\frac{{\phi _{n - k} h^{ - k} }}{{\left( {s - 1/h} \right)^{k + 1} }}} 
\label{equ22}
\end{equation}

In the same way, we can handle eq.(\ref{equ16}), but, in this case, by applying more conveniently a backward Euler scheme because of the different boundary condition. Calling ${\bf{Z}} = \left[ {\begin{array}{*{20}c}
   {z_{n} } & {z_{n + 1} } & {...}  \\ \end{array}} \right]^T $, $
{\bf{\Phi }} = \left[ {\begin{array}{*{20}c}
   {\phi _{n } } & {\phi _{n + 1} } & {...}  \\
\end{array}} \right]^T $ it is easy to show that the resolvent equation can be formally expressed in matrix notation as
\begin{equation}
{\bf{Z}} = \left( {{\bf{I}} - \left( {1 - sh} \right)^{ - 1} {\bf{A}}} \right)^{ - 1} {\bf{\Phi }}h\left( {1 - sh} \right)^{ - 1} 
\label{equ23}
\end{equation}
and that the general term $z_n$ is found to be
\begin{equation}
z_n  = \sum\limits_{k = 0}^\infty  {\left( { - 1} \right)^{k+1} \frac{{\phi _{n + k} h^{ - k} }}{{\left( {s - 1/h} \right)^{k + 1} }}} 
\label{equ24}
\end{equation}

Introducing the expression found for $y_n$ and $z_n$ into eq.(\ref{equ17}) and performing the integral inside the sum, one has to evaluate the two integrals 
\begin{subequations}
\begin{eqnarray}\label{equ25}
\frac{1}{{2\pi i}}\int_\Gamma ^{} {\frac{{F_-\left( s \right)}}{{\left( {s - 1/h} \right)^{k + 1} }}ds = \frac{1}{{k!}}\left. {\frac{{d^k F_-\left( s \right)}}{{ds^k }}} \right|_{s =   1/h} }  =\bar{\alpha}  _k(f) 
\label{equ25a}
\end{eqnarray}
\begin{eqnarray}
\frac{1}{{2\pi i}}\int_\Gamma ^{} {\frac{{F_+\left( s \right)}}{{\left( {s - 1/h} \right)^{k + 1} }}ds = \frac{1}{{k!}}\left. {\frac{{d^k F_+\left( s \right)}}{{ds^k }}} \right|_{s = 1/h} }  = \bar{\omega} _k(f) 
\label{equ25b}
\end{eqnarray}
\end{subequations}
due to the Cauchy's integral formula for the derivatives. 
The notation adopted to indicate the coefficients $\bar{\alpha}_k(f)$ and $\bar{\omega}_k(f)$ is meant to underline their dependence on the function $f$.

Summing up, from the non-linear function $f(x)$ we obtain $\hat{f}$ performing the inverse Fourier transform and by eqs.(\ref{eq12}) the functions $F_\pm(s)$ are calculated. 
These functions are expressed by Taylor expansion around $s=1/h$ and the coefficients are $\bar{\alpha}_k$ and $\bar{\omega}_k$. 
Once the coefficients are known, (\ref{equ17}) is finally expressed in the form
\begin{equation}
  \left( {T_f \phi } \right)\left( \theta  \right) = E\left[ {f\left( X \right)e^{i\theta X} } \right] = \mathop {\lim }\limits_{h \to 0^ +  } \left( {\sum\limits_{k = 0}^\infty  {\alpha _k(f) \phi \left( {\theta  - kh} \right)}  + \sum\limits_{k = 0}^\infty  {\omega _k(f) \phi \left( {\theta  + kh} \right)} } \right)
\label{equ26}
\end{equation}
having indicated with $\alpha _k(f)  = \left( { - 1} \right)^{k + 1} h^{ - k} \bar \alpha _k(f) $ and with $\omega _k (f) = \left( { - 1} \right)^{k + 1} h^{ - k} \bar \omega _k (f)$.

If one selects a fixed small value of the increment $h$ in the latter expression, it corresponds to a first order approximation of the operator $(T_f\phi)(\theta)$, otherwise, the expression is exact to the limit.

\subsection{Examples}
Some simple examples will show how to practically apply the relations found. In this section, for readability's sake, the dependence of the coefficients on the function will not be indicated because evident from the contest.

\subsubsection{SDE with linear drift.}

The representation formula of the non-linear drift given in eq.(\ref{equ26}) is valid under the hypothesis that the involved integral transforms exist, at least in the generalized sense. Let us first consider a simple linear system excited by external Gaussian noise, that is with $a(X)=a=$ constant and $b(X)=1$
\begin{equation}
dX\left( t \right) =  - a X\left( t \right) + dB\left( t \right),\,\,\,\,\,\,\,\,\,\,\,\,\,\,\,\,\,\,a > 0
\label{eqE1}
\end{equation}
The SFPE equation in this case is 
\begin{equation}
\frac{{\partial \phi }}{{\partial t}} =  - i a\theta E\left[ {Xe^{i\theta X} } \right] - \frac{{\theta ^2 }}{2}\phi 
\label{eqE2}
\end{equation}
which becomes a partial differential equation once is recognized that $E\left[ {Xe^{i\theta X} } \right] =  - i\frac{{\partial \phi }}{{\partial \theta }}$, as stated in eq.(\ref{eq_3}). We follow the procedure proposed to attain to this simple result.

In this case, $f(x)=-ax$, and the inverse Fourier transform is $\hat{f}(\theta)= -{i} a\, \delta'(\theta)$, where $\delta$ is the Dirac's delta and the prime indicates the first derivative, and $F_\pm(s)=\mp {i\, a}\, s/2$ by direct integration.
The integrals in the Laplace transforms have been performed with lower limit in $0$, and not in $0-$, and for this reason the factor $1/2$ appears. 
It is easy to find the Taylor series in $s=1/h$ for both functions $F_\pm(s)$
\begin{eqnarray}
 F_ +  \left( z \right) =  - \frac{{i a}}{{2h}} - \frac{{i a}}{{2}}\left( {z - \frac{1}{h}} \right)  \nonumber \\
 F_ -  \left( z \right) = \frac{{i a}}{{2h}} + \frac{{i a}}{{2}}\left( {z - \frac{1}{h}} \right)  \nonumber 
\end{eqnarray}
and consequently to calculate the coefficients that are
\begin{eqnarray}
 \bar{\alpha} _0  =  - \frac{{ia}}{{2h}};\,\,\,\bar{\alpha} _1  =  - \frac{{ia}}{2};\,\,\,\bar{\alpha} _2  = \bar{\alpha} _3  = ... = 0 \nonumber \\ 
 \bar{\omega} _0  = \frac{{ia}}{{2h}};\,\,\,\bar{\omega}_1  = \frac{{ia}}{2};\,\,\,\bar{\omega} _2  = \bar{\omega} _3  = ... = 0 \nonumber
\end{eqnarray}

Lastly, introducing the latter coefficient in eq.(\ref{equ26}) and making the limit $h \to 0$  we obtain
\begin{eqnarray}
\mathop {\lim }\limits_{h \to 0} \left( { - \frac{{ia}}{{2h}}\left( {\phi \left( \theta  \right) - \phi \left( {\theta  - h} \right)} \right) + \frac{{ia}}{{2h}}\left( {\phi \left( \theta  \right) - \phi \left( {\theta  + h} \right)} \right)} \right) \to  - i a\,\frac{{\partial \phi }}{{\partial \theta }}
\nonumber
\label{equ27}
\end{eqnarray} 
that is the searched result. Without taking the limit, the construction of the operator $(T_f\phi)(\theta)$ produces the first order approximation of $\partial \phi /\partial \theta $ by finite differences.

\subsubsection{SDE with non-linear polynomial drift.}

Generalization to non-linear drift term of power-law type, i.e. $f(x)=x^j$, with $j \in \mathbb{N}$ follows plainly. Indeed, in this case $\hat{f}(\theta)={i}^j \, \delta^{(j)}(\theta)$, where the apex between brackets indicates the jth derivative and $F_\pm(s)=(\pm {i})^j s^j$. Substitution in eq.(\ref{equ25a}) and (\ref{equ25b}) gives
\[
\bar{\alpha} _k  = \frac{{i}}{{\rm{2}}}^j h^{k - j} \left( {\begin{array}{*{20}c}
   j  \\
   k  \\
\end{array}} \right);\,\,\,\,\,\,\,\,\,\,\bar{\omega} _k  = \frac{{\left( {{ - i}} \right)^j }}{2}h^{k - j} \left( \begin{array}{l}
 j \\ 
 k \\ 
 \end{array} \right)
\]
that, introduced in (\ref{equ26}) gives $ (T_f\phi)(\theta) = \left({-{i}} \right)^k \phi ^{\left( k \right)} \left( \theta  \right)$, for $h \to 0$.

\subsubsection{SDE with real power-law non-linear drift.}

Consider a drift term of the form $ f\left( x \right) = \left| x \right|^\gamma  {\mathop{\rm sgn}} \left( x \right)$. 
Non-linearity of this kind is frequently used to model viscous-elasticity. 
L\'evy driven non-linear system with this kind of drift have been studied for example in \cite{chec04} and \cite{cott10b} by applying the fractional calculus.
Application of the procedure proposed leads to interesting results, in accordance with the above cited papers. 

First of all, by means of Mathematica, it is found that 
\[
  \hat{f}\left( \theta  \right) =  - \frac{{i \cos \left( { \gamma \pi {\rm{/2}}} \right)\Gamma \left( {1 + \gamma } \right)}}{\pi }\left| \theta  \right|^{ - 1 - \gamma } {\rm sgn} \left( \theta  \right);\,\,\,\,\,\,\,\,F_ \pm  \left( s \right) =  \pm \frac{{i}\,s^\gamma  }{{2\sin \left( { \gamma \pi/2} \right)}}
\]
The coefficients 
\[
\bar{\alpha} _k  = \frac{i}{{{\rm{2 sin}}\left( {\gamma \pi {\rm{/2}}} \right)}}h^{\gamma  - k} \left( {\begin{array}{*{20}c}
   \gamma   \\
   k  \\
\end{array}} \right);\;\;\;\;\;\;\bar{\omega} _k  =  - \frac{{i}}{{{2 \sin}\left( {\gamma \pi {\rm{/2}}} \right)}}h^{\gamma  - k} \left( {\begin{array}{*{20}c}
   \gamma   \\
   k  \\
\end{array}} \right);
\]
introduced in (\ref{equ26}) give that the integral $(T_f\phi)(\theta)$ coincides in this case with a fractional operator of potential type (see \cite{samko}, p. 214), that is
\[
(T_f\phi)(\theta)\equiv \left( {H^{ - \gamma } \phi } \right)\left( \theta  \right) = \frac{1}{{2\sin \left( {\gamma \pi /2} \right)}}\left( {\left( {D_ + ^\gamma  \phi } \right)\left( \theta  \right) - \left( {D_ - ^\gamma  \phi } \right)\left( \theta  \right)} \right)
\]
where $\left( {H^{ - \gamma } \phi } \right)\left( \theta  \right)$ is called complementary Riesz fractional derivative of real order $\gamma>0$, $\gamma \ne 2,4,...$. It can also be defined in terms of Gr\"unwald-Letnikov fractional derivatives in the form
\begin{equation}
\left( {D_ \pm ^\gamma  \phi } \right)\left( \theta  \right) = \lim _{h \to 0} h^{ - \gamma } \sum\limits_{k = 0}^\infty  {\left( { - 1} \right)^k \left( \begin{array}{l}
 \gamma  \\ 
 k \\ 
 \end{array} \right)\phi \left( {\theta  \mp kh} \right)} 
\label{eqGL}
\end{equation}
and such coefficient coincides with those found by the proposed convolution quadrature.

\textbf{Remark:} The previous examples show the relations between the transform operator $(T_f\phi)(\theta)$ and the classical derivatives and the fractional derivatives. 

The first two examples showed that $(T_f\phi)(\theta)$ coincides with the classical derivatives of integer order $k$, with $k \in \mathbb{Z}$ in case the function $f(x)=x^k$ is selected.

If we consider $f\left( x \right) = \left| x \right|^{ - \gamma } $ or $f\left( x \right) = \left| x \right|^{ - \gamma } {\rm{sgn}}(x) $, with  $\gamma \in \mathbb{C}$, then the third example shows that the integral transform $(T_f\phi)(\theta)$ coincides with a particular Riesz fractional integral. Moreover, the numerical procedure based on the convolution quadrature through the Eulerian schemes to solve (\ref{equ15}) and (\ref{equ16}) gives: 1) the finite difference schemes of first order, in case of integer power-law kernel; 2) the Gr\"unwald-Letnikov fractional approximation in case of complex power-law kernel.

This similarity suggests that many other properties typical of fractional calculus, such as the integration by part formula, the composition rules and the definition on a bounded interval, may be stated also for the integral $(T_f\phi)(\theta)$.
We report such properties in Appendix A, for readability's sake. Conditions for the convergence of the sum are reported in Appendix B. 

\section{Solution of scalar non-linear stochastic systems}
Let us revert to the statistic of the solution of the non-linear systems (\ref{eq_1}). 
Introducing the transform $(T_f\phi)(\theta)$ in (\ref{eq_2}) the integro-differential equations in terms of the characteristic function $\phi(\theta,t)$ can be written in the form
\begin{equation}
\frac{{\partial \phi \left( {\theta ,t} \right)}}{{\partial t}} = i\theta \left( {T_f \phi } \right)\left( \theta  \right) - g\left( \theta  \right)\phi \left( {\theta ,t} \right)
\label{equ30}
\end{equation}
in which $g(\theta)$ is equal to $\theta^2/2$, $\lambda(1-\phi_Y(\theta))$, $\left| \theta  \right|^\alpha  $ in the case of Gaussian, Poisson and $\alpha$-stable external white noise, respectively.
To solve this equation, let us select a finite interval $[-{\bar \theta },{\bar \theta }]$ and divide it in equidistant subinterval of finite dimension $h={\bar \theta }/m$, with $m \in \mathbb{N}$.
The choice of a  symmetric interval depends on the symmetry of the characteristic function.
Moreover, indicate by $\theta _j  = jh$, $\phi _j  = \phi \left( {\theta _j } \right)$, with $j =  - m,...,m$.
Formula to express $(T_f\phi)(\theta)$ in the bounded domain $[-{\bar \theta },{\bar \theta }]$ is given in Appendix, eq.(\ref{equ29}).
Then (\ref{equ30}) is rewritten as
\begin{eqnarray}
  \frac{{\partial \phi _j }}{{\partial t}} = i\theta _j \left( {\sum\limits_{k = 0}^{\left[\frac{{\theta _j  + \bar \theta }}{h}\right]} {\alpha _k \left( f \right)\phi _{j - k}  + \sum\limits_{k = 0}^{\left[\frac{{\bar \theta  - \theta _j }}{h}\right]} {\omega _k \left( f \right)\phi _{j + k} } } } \right) - \phi _j g\left( {\theta _j } \right)
\label{equ30a}
\end{eqnarray}

As the latter equation is valid for every $j=-m,...,m$, the linear system of differential equations with the form
\begin{equation}
{\boldsymbol {\dot \phi}}  = {\bf{R}}{\boldsymbol {\phi}}
\label{equ31}
\end{equation}
can be constructed, in which ${\bf{R}} = {i \bf{UT_f}} - {\bf{G}}$,
\begin{equation}
{\bf{U}} = \left\lceil { - \theta _m ,...,0,...,\theta _m } \right\rfloor 
\label{equ32}
\end{equation}
is a diagonal matrix,
\begin{equation}
  {\bf{T}}_{\bf{f}}  = \left( {\begin{array}{*{20}c}
   {\alpha _0 \left( f \right) + \omega _0 \left( f \right)} & {\omega _1 \left( f \right)} & {\omega _2 \left( f \right)} & {} & {\omega _{2m} \left( f \right)}  \\
   {\alpha _1 \left( f \right)} & {\alpha _0 \left( f \right) + \omega _0 \left( f \right)} & {\omega _1 \left( f \right)} &  \cdots  & {\omega _{2m - 1} \left( f \right)}  \\
   {\alpha _2 \left( f \right)} & {\alpha _1 \left( f \right)} & {\alpha _0 \left( f \right) + \omega _0 \left( f \right)} &  \ddots  & {}  \\
   {} &  \vdots  &  \ddots  &  \ddots  & {\omega _1 \left( f \right) }  \\
   {\alpha _{2m} \left( f \right)} & {\alpha _{2m - 1} \left( f \right)} & {} & {\alpha _1 \left( f \right)} & {\alpha _0 \left( f \right) + \omega _0 \left( f \right)}  \\
\end{array}} \right)
\label{equ33}
\end{equation}
is a Toeplitz matrix of coefficients depending on the non-linearity, and ${\boldsymbol\phi} ^T  = \left[ {\phi _{ - m} ,...,\phi _m } \right]^T $ is the vector collecting the $2m+1$ unknown values of the characteristic function at the grid points. 

The matrix ${\bf{G}}$ is 
\begin{enumerate}
	\item ${\bf{G}} = 1/2\left\lceil { - \theta^2 _m ,...,0,...,\theta^2 _m } \right\rfloor$ for Gaussian noise;
	\item ${\bf{G}} = \lambda \left\lceil { 1-\phi_Y(-\theta_m) ,...,0,...,1-\phi_Y(\theta_m)} \right\rfloor$ for Poisson noise;
	\item ${\bf{G}} = \lambda \left\lceil { \left|- \theta_m  \right|^\alpha  ,...,0,..., \left| \theta_m  \right|^\alpha} \right\rfloor$ for $\alpha$-stable L\'evy noise.
\end{enumerate}

In the stationary case $\dot{\boldsymbol{\phi}}={\bf{0}}$, and the system of differential equations (\ref{equ31}) becomes an algebraic linear system, that can be solved, by posing the relevant boundary condition $\phi_0=\phi(0)=1$. 
Such a boundary condition can be easily posed by defining the reduced $2m$ dimensional vector of coefficients ${\bf{b}}^T  = \left[ {\begin{array}{*{20}c}
   { \omega _m } & {...} & { \omega _1 } & {\alpha _1 } & {...} & {\alpha _m }  \\
\end{array}} \right]^T $
and the reduced matrix $\bf{\bar{R}}$ obtained by deleting both the central line and the central row to the matrix $\bf{{R}}$. In this way the reduced vector of unknown ${\boldsymbol{\bar{\phi}}}^T  = \left[ {\phi _{ - m} ,...,\phi _{ - 1} ,\phi _1 ...\phi _m } \right]^T $ is the solution of the simple linear system in the form
\begin{equation}
{\boldsymbol{\bar{\phi}}} = - {\bf{\bar R}}^{ - 1} \,\,{\bf{\bar b}}
\label{equ34}
\end{equation}

The non-stationary solution can be found by solving the linear system of differential equations (\ref{equ31}). This can be simply achieved by spectral decomposition.

\section{Validation and Applications}

To assess the accuracy of the proposed procedure we start studying the SDE in the It\^o's form
\begin{equation}
dX\left( t \right) = \left( { - X \left( t \right) + \cos \left( {X\left( t \right)} \right)} \right)dt + dB\left( t \right)
\label{eqV1}
\end{equation}
in which $dB(t)$ is the standard Brownian motion, with $q=1$.
In the stationary case, the statistic of the solution are known in analytical form and they can be used to validate the proposed method.
In fact the stationary density is given as 
\[
p_s \left( x \right) = C\,e^{2\int_0^x {\left( { - \xi  + \cos \xi } \right)d\xi } } 
\]
being $C$ a normalization constant (see \cite{sobc91}).

For this application we selected $\bar{\theta} = 15$, and different mesh sizes $m = 100; 150; 200; 500$ to investigate on the convergence of the numerical approximation to the exact solution and the influence of $h$. 
The linear algebraic system of equations given in (\ref{equ34}), which returns the stationary characteristic function, has been solved and then by inverse numerical inverse Fourier transform, the stationary density has been found for every value of $h$ and it is indicated as $p_h(x)$.
 
\begin{figure}[t]
\centering
\includegraphics[width = 8cm]{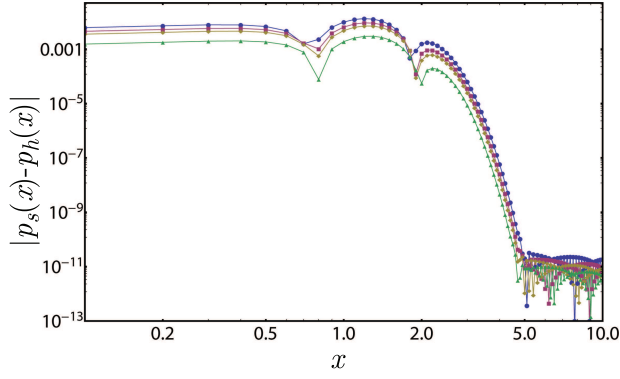}
\caption{Absolute error for different values of $h$: $h=0.15$ circles, $h=0.10$ squares, $h=0.075$ rhomboidal and $h=0.030$ triangles}
\label{Figure1}
\end{figure}

The coefficients defining the numerical form of the integral transform $(T_f\phi)(\theta)$ are
\[
\alpha _j  = \frac{{h^{ - j} }}{2}\left( { - \frac{{e^{ - 1/h} }}{{\Gamma \left( {1 + j} \right)}} - i\frac{{\left( { - 1} \right)^{ - j} }}{h}r_j } \right)
\]
and 
\[
\omega _j  = \frac{{h^{ - j} }}{2}\left( { - \frac{{e^{ - 1/h} }}{{\Gamma \left( {1 + j} \right)}} + i\frac{{\left( { - 1} \right)^{ - j} }}{h}r_j } \right)
\]
where $r_j =1$ for $j=0$ and $r_j=h$ for $j\ne0$. 
Figure (\ref{Figure1}) shows the absolute error, calculated in the positive axis, between the exact and the approximate stationary density of the solution $p_h(x)$.
This result shows that by this method it is possible to achieve a good numerical approximation keeping an absolute error that is at most of order $h$. 
This order of magnitude derives from having used an Eulerian scheme in finding the series approximation of the integral transform $(T_f\phi)(\theta)$. 
In order to reduce such an error, higher order approximations can be used and their implementation is underway by the author.

\begin{figure}[t]
 \centering
 \subfigure[]
   {\includegraphics[width=6.5cm]{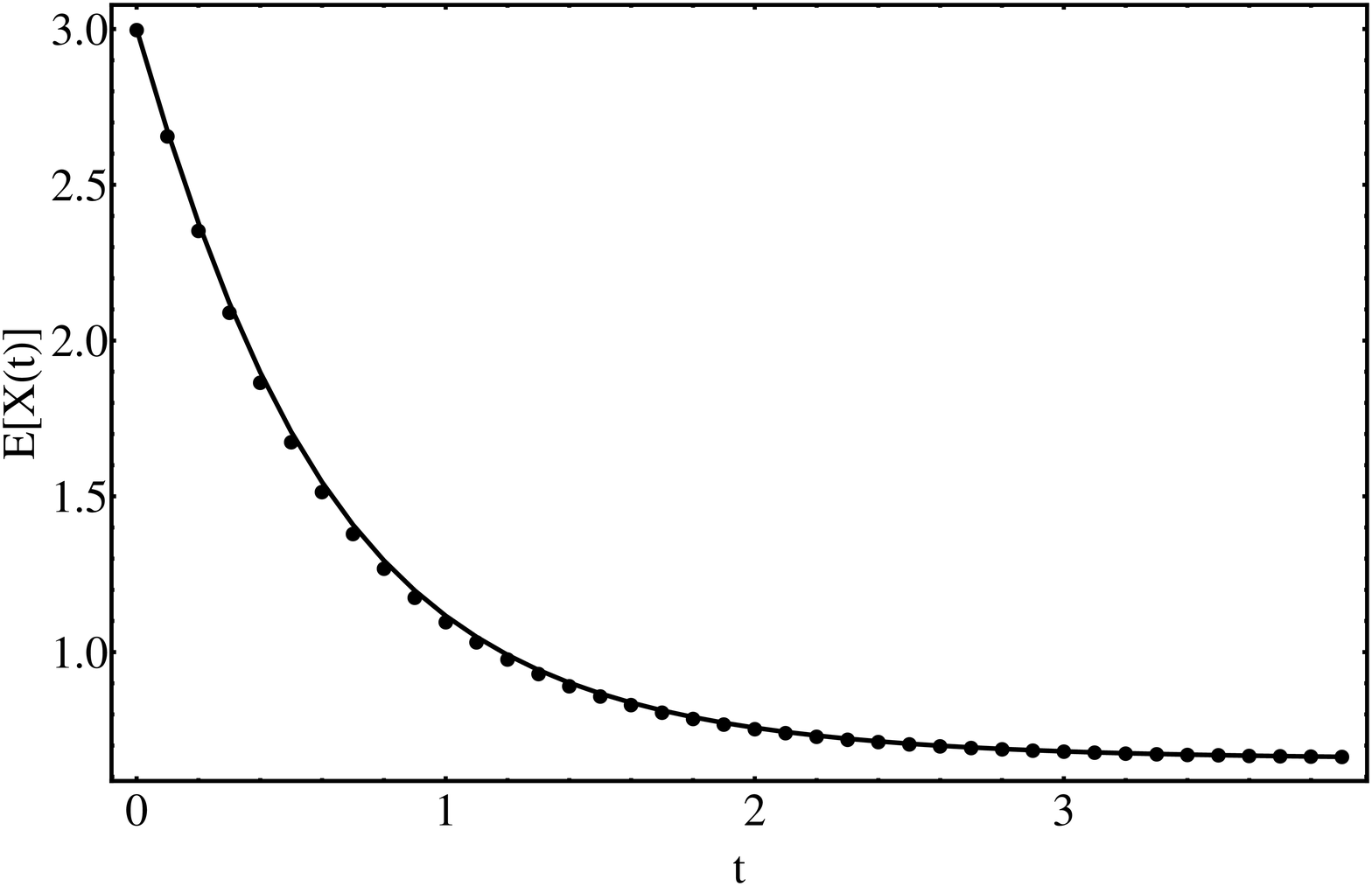}}
 \subfigure[]
   {\includegraphics[width=6.5cm]{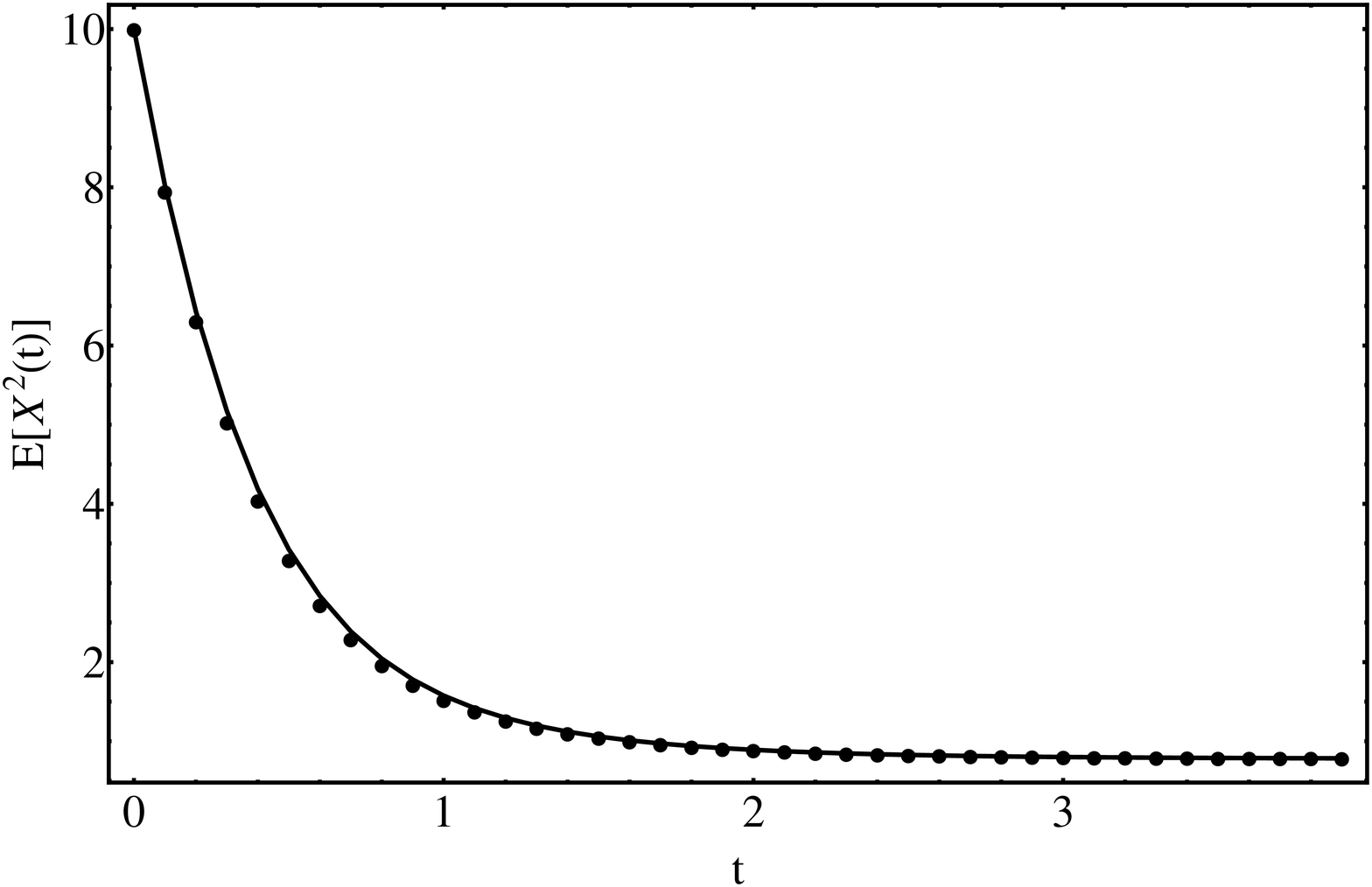}}
  \caption{Time evolution of the first and second order moments. Comparison between the results calculated from the eigenfunction expansion of the FPK (continuous line) and the proposed method (dotted)} 
 \label{Figure2}
 \end{figure}

The non-stationary density of this system has also been calculated for $m=100$ by solving eq.(\ref{equ31}). For this case of external Gaussian excitation, the solution of the Fokker-Planck equation can be approximated also by eigenfunction expansion (see for instance \cite{risk96}, p.101). In Figure (\ref{Figure2}) a comparison in term of first and second order moments calculated from the non-stationary densities calculated by the proposed method and the eigenfunction expansion method are compared, showing good agreement.

After having validated the method, in the following, we consider the non-stationary solution of the three non-linear SDEs:
\[
\left\{ \begin{array}{l}
 dX\left( t \right) = \left( { - X\left( t \right) + \cos \left( {X\left( t \right)} \right)} \right)dt + dL\left( t \right) \\ 
 X\left( 0 \right) = X_0  \\ 
 \end{array} \right.
\]
in which: 

1) for $dL(t) = dB(t)$, the initial condition is Gaussian. i.e. $X_0 = N\left({3,1} \right)$ and $m=100$; 

2) for  $dL(t) = dC(t)$ we consider exponential distributed jumps with mean $\mu_Y = 3$, $\lambda_Y =3$, Gaussian distributed initial condition $X_0 = N\left({0,1} \right)$ and $m=100$; 

3) for $dL(t) = dL_{\alpha}(t)$, i.e. in case of $\alpha$-stable L\'evy excitation, we selected $\alpha = 3/2$, uniform initial condition $X_0 = U(0,1)$ and $m=100$.

It is usefull to recall that the Fokker-Planck equations, governing the densities, for these systems have the forms
\begin{subequations}\label{eq33} 
\begin{eqnarray}
\frac{{\partial p}}{{\partial t}} =  - \frac{\partial }{{\partial x}}\left( {\left( { - x + \cos \left( x \right)} \right)p} \right) + \frac{1}{2}\frac{{\partial ^2 p}}{{\partial x^2 }}
\label{eq33a}  
\end{eqnarray}
\begin{eqnarray}
\frac{{\partial p}}{{\partial t}} =  - \frac{\partial }{{\partial x}}\left( {\left( { - x + \cos \left( x \right)} \right)p} \right) - \lambda p + \lambda \int_{ - \infty }^\infty  {p\left( \xi  \right)p_Y \left( {x - \xi } \right)d\xi } 
\label{eq33b}
\end{eqnarray}
\begin{eqnarray}
\frac{{\partial p}}{{\partial t}} =  - \frac{\partial }{{\partial x}}\left( {\left( { - x + \cos \left( x \right)} \right)p} \right) + \left( {\cal{D}}^\alpha p \right)\left( x \right)
\label{eq33c}
\end{eqnarray}
\end{subequations}
in which $\cal{D}^\alpha$ is the Riesz fractional derivative, see \cite{metz00} for the derivation of the equation and \cite{samko} for relevant properties of this operator.
The numerical solution of the latter equations is not trivial: In the case of Gaussian excitation, in which the eq.(\ref{eq33a}) is a partial differential equation, numerical evaluation of the eigenfunctions and eigenvalues of the Fokker-Planck operator for obtaining the non-stationary solution is required (see \cite{risk96}, ch. 5); 
Eqs.(\ref{eq33b}) and (\ref{eq33c}) are respectively a partial integro-differential and a partial fractional differential equations and quadrature methods can be applied. 

\begin{figure}[t]
 \centering
 \subfigure[]
   {\includegraphics[width=6.5cm]{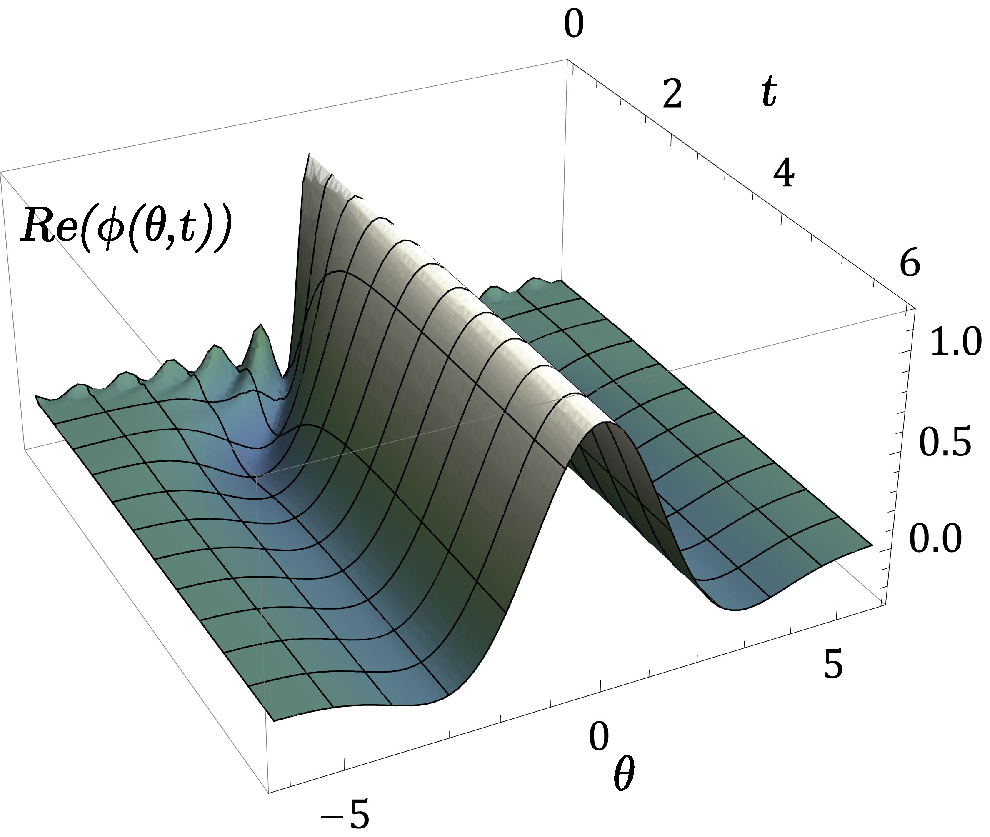}}
 \subfigure[]
   {\includegraphics[width=6.5cm]{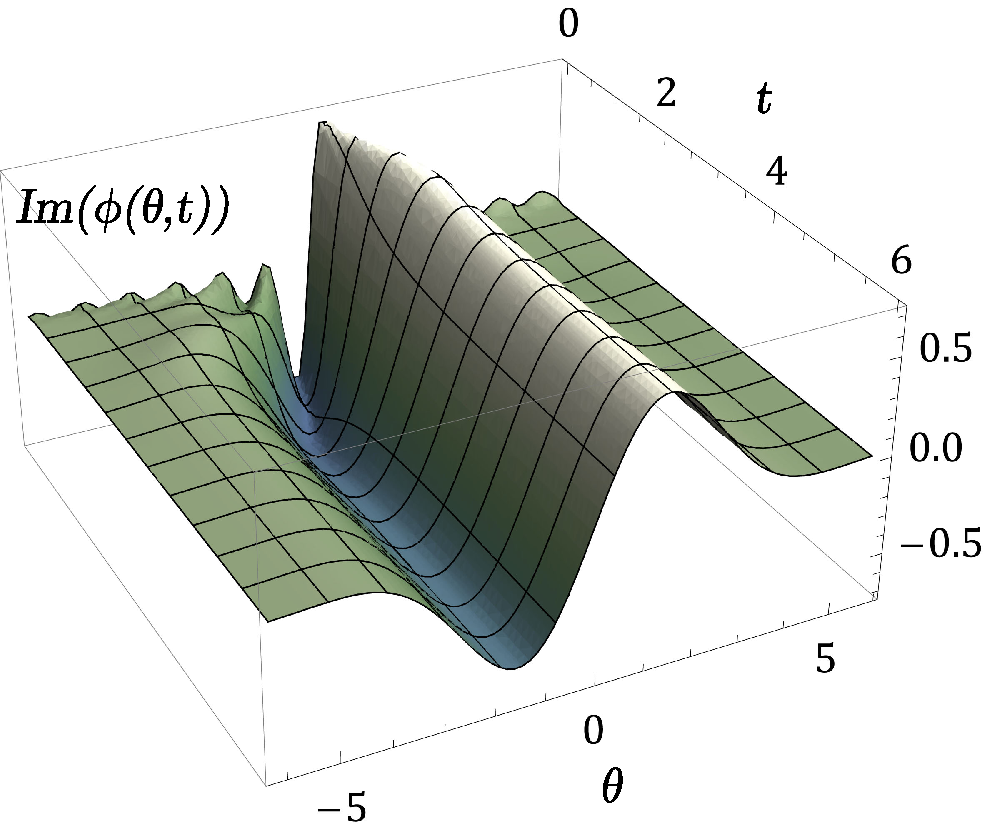}}
 \subfigure[]
   {\includegraphics[width=6.5cm]{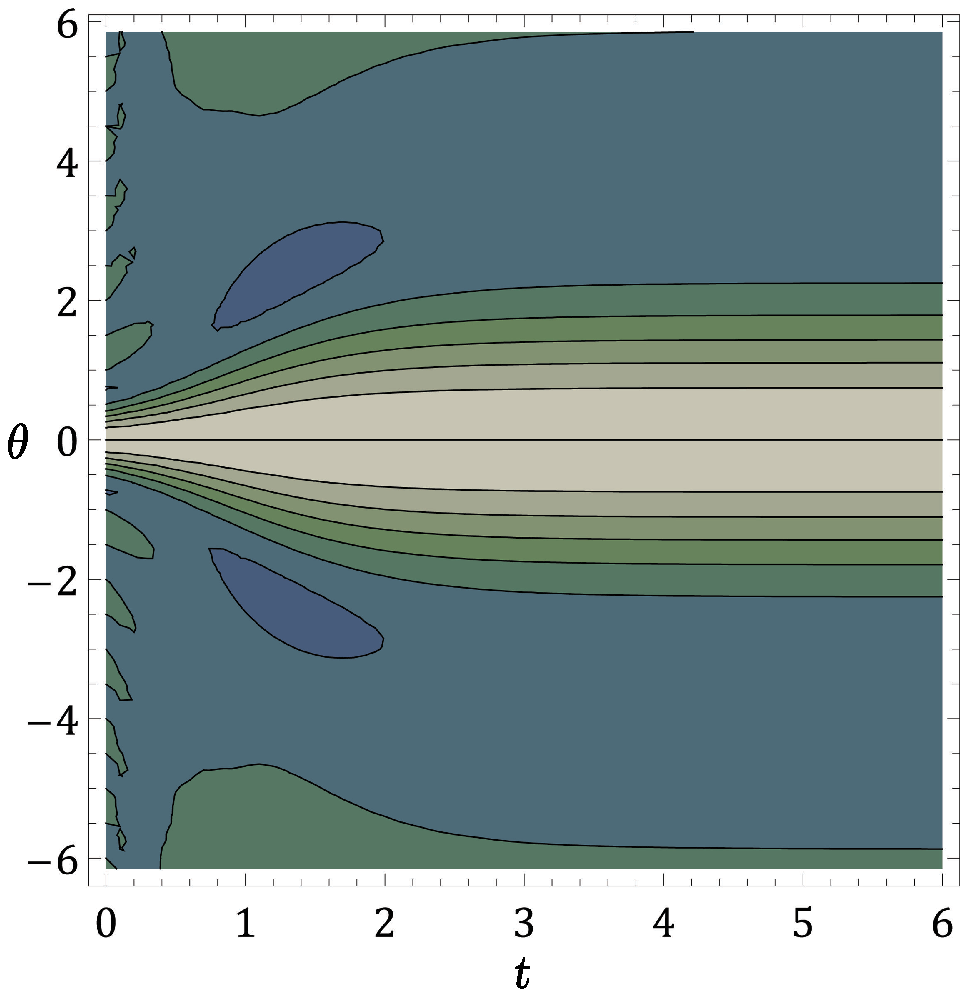}}
 \subfigure[]
   {\includegraphics[width=6.5cm]{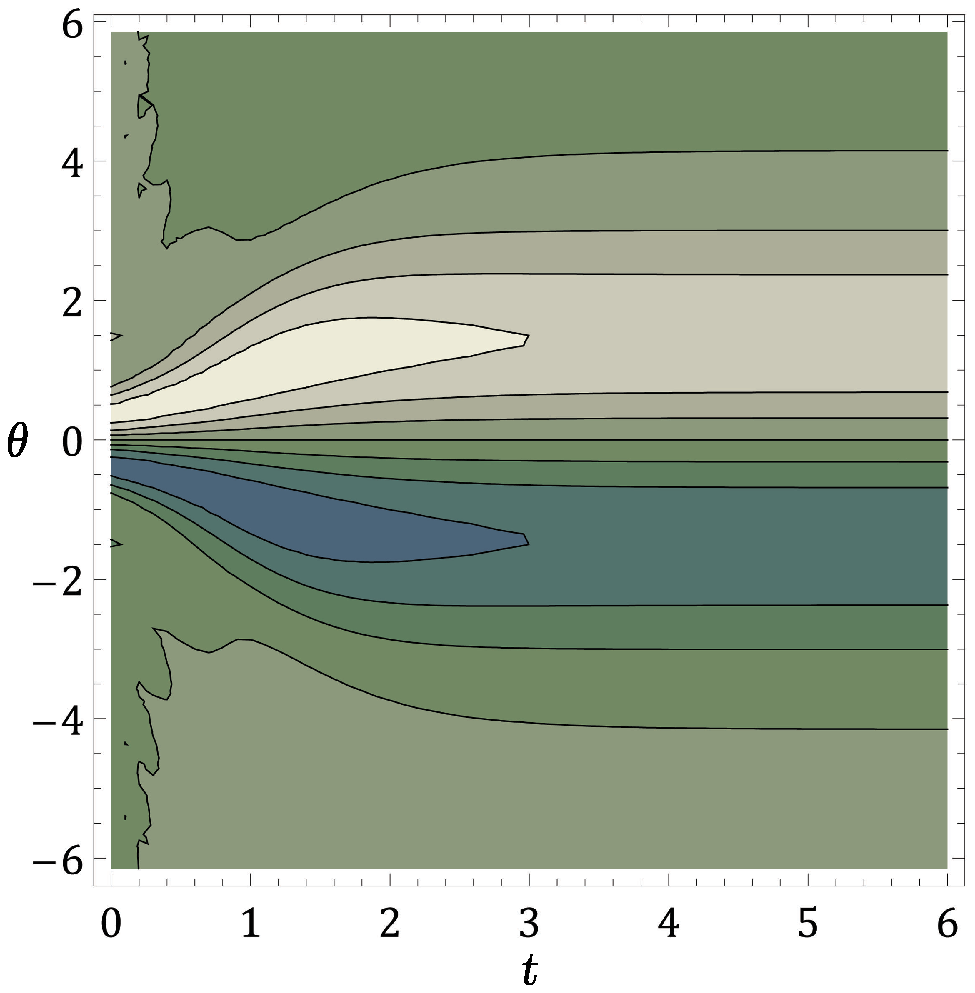}}
 \caption{Gaussian white noise excitation: a) Real part of the CF; b) Imaginary part of the CF; c) Contour plot of $Re(\phi(\theta,t))$; d) Contour plot of $Im(\phi(\theta,t))$}
 \label{Figure3}
 \end{figure}

\begin{figure}[t]
 \centering
 \subfigure[]
   {\includegraphics[width=6.5cm]{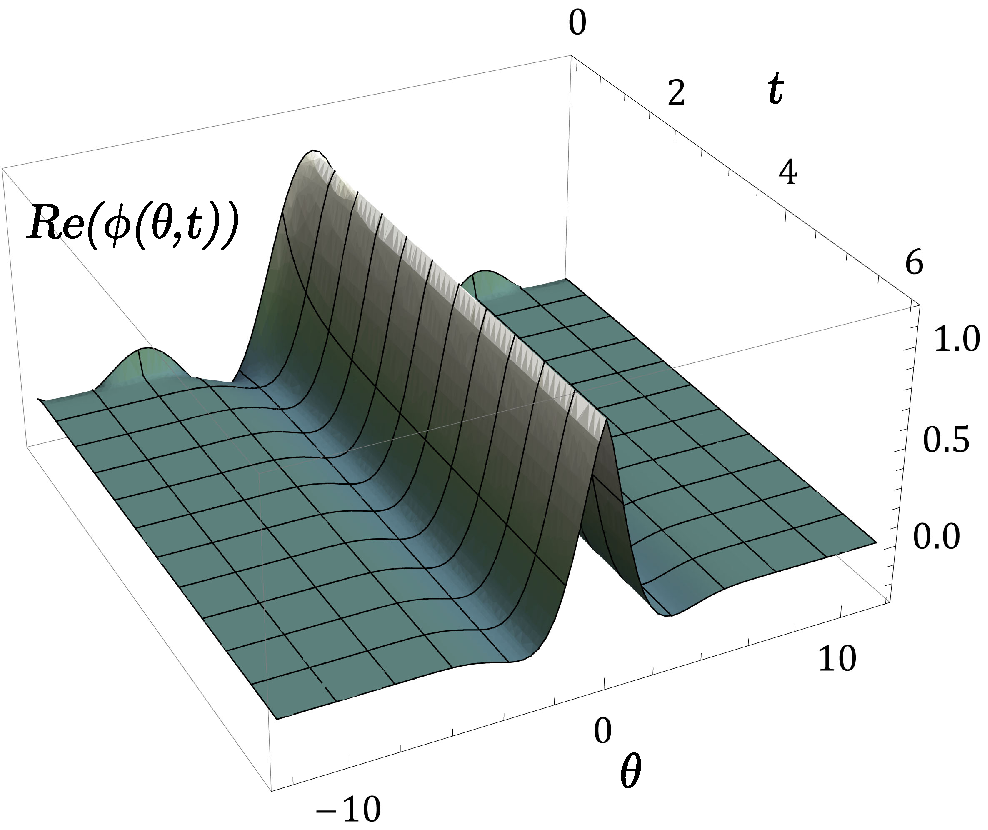}}
 \subfigure[]
   {\includegraphics[width=6.5cm]{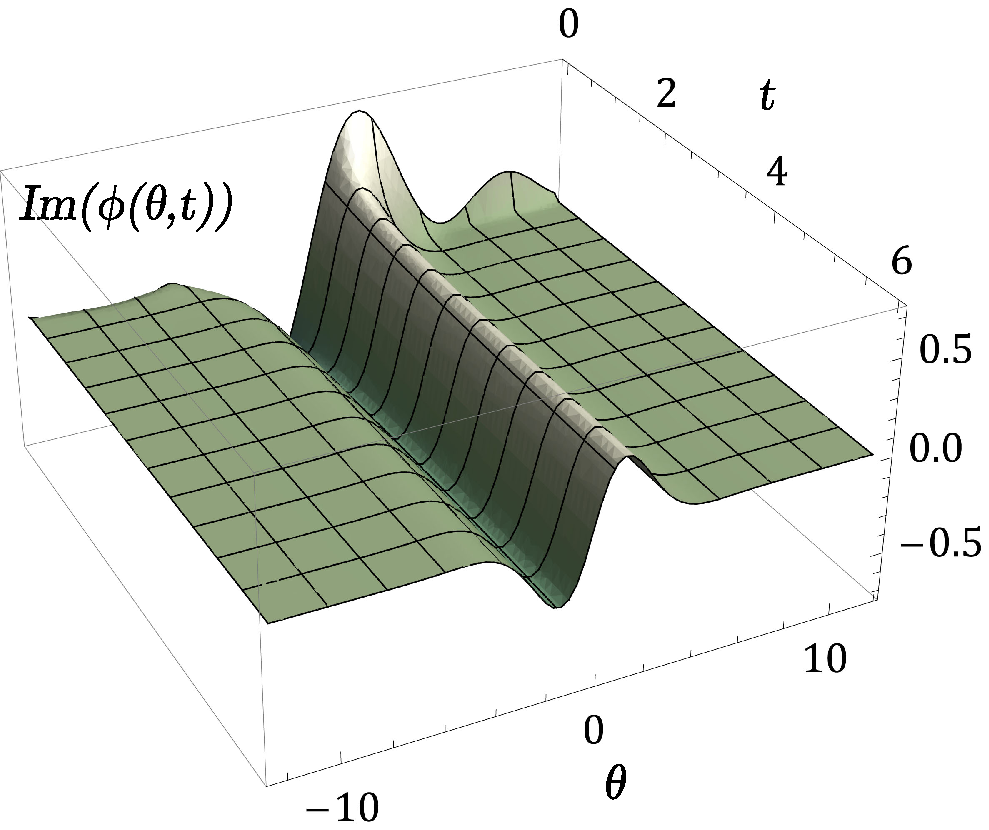}}
 \subfigure[]
   {\includegraphics[width=6.5cm]{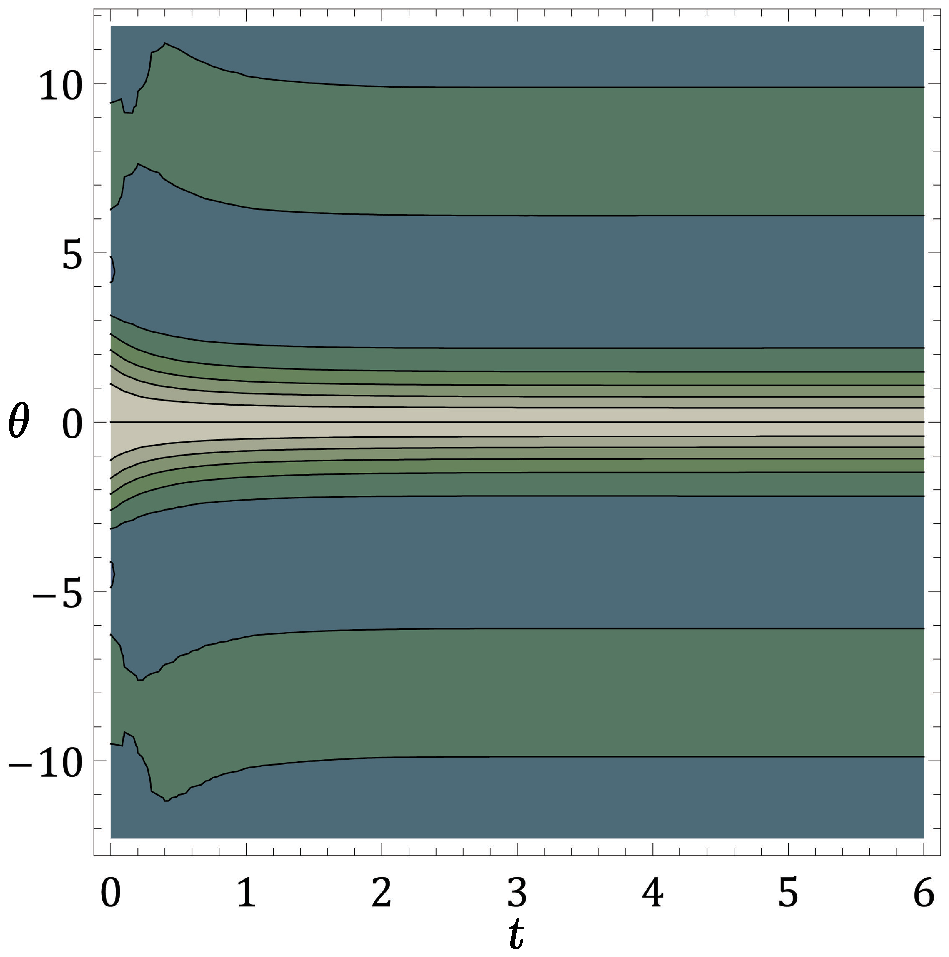}}
 \subfigure[]
   {\includegraphics[width=6.5cm]{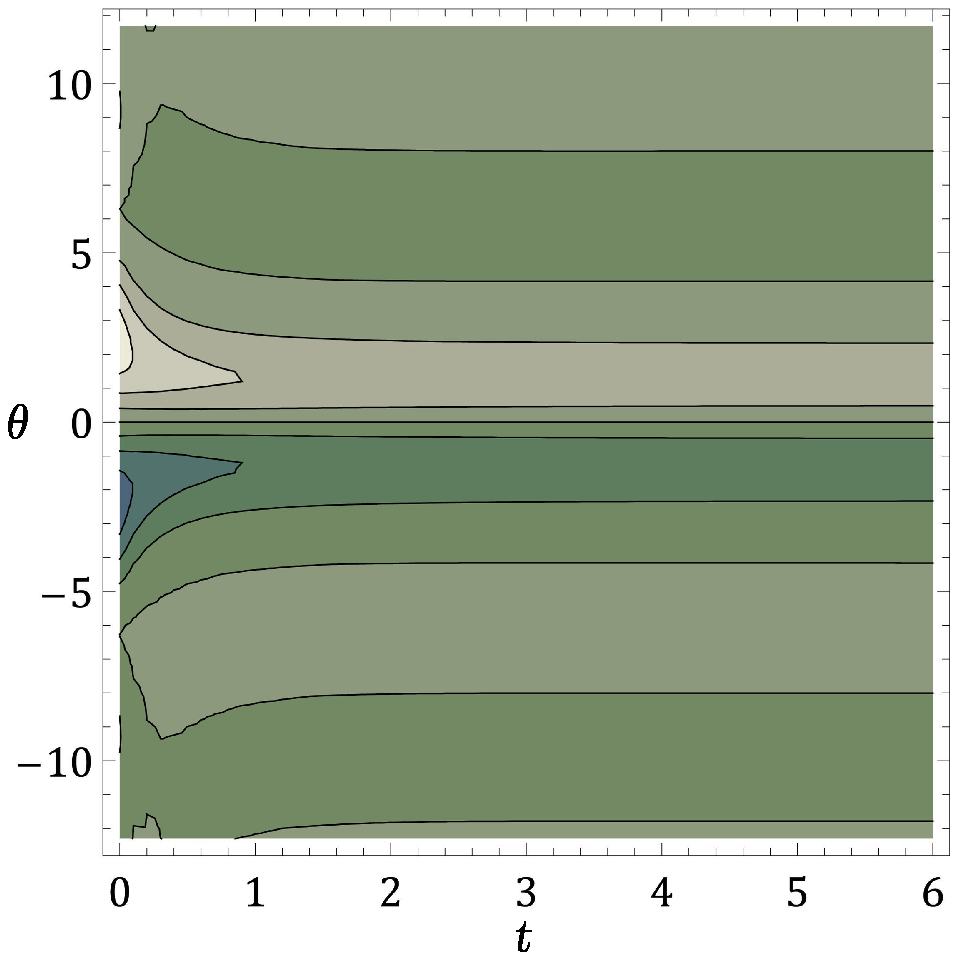}}
 \caption{Poisson white noise excitation: a) Real part of the CF; b) Imaginary part of the CF; c) Contour plot of $Re(\phi(\theta,t))$; d) Contour plot of $Im(\phi(\theta,t))$}
 \label{Figure4}
 \end{figure}

\begin{figure}[t]
 \centering
 \subfigure[]
   {\includegraphics[width=6.5cm]{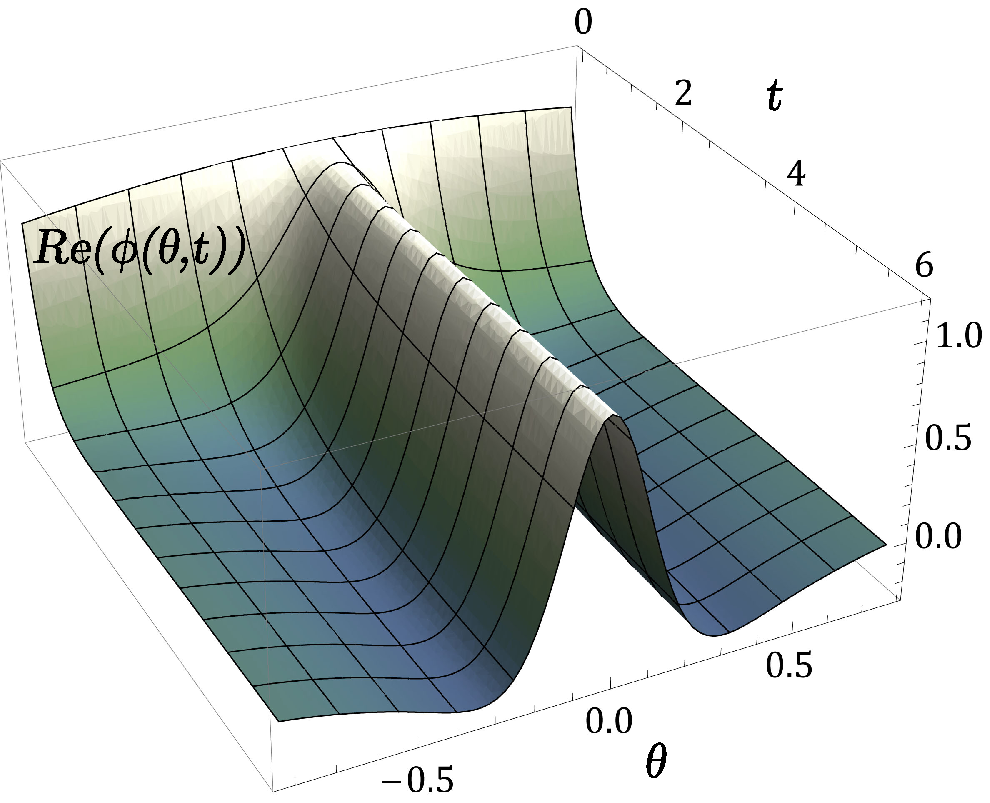}}
 \subfigure[]
   {\includegraphics[width=6.5cm]{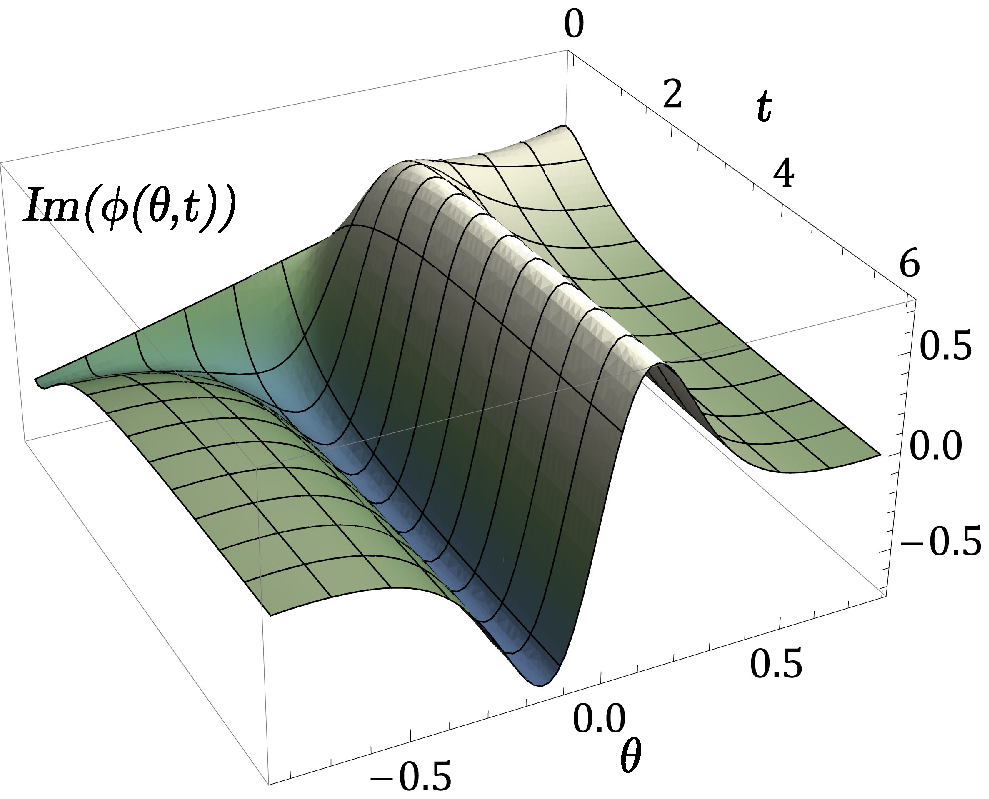}}
 \subfigure[]
   {\includegraphics[width=6.5cm]{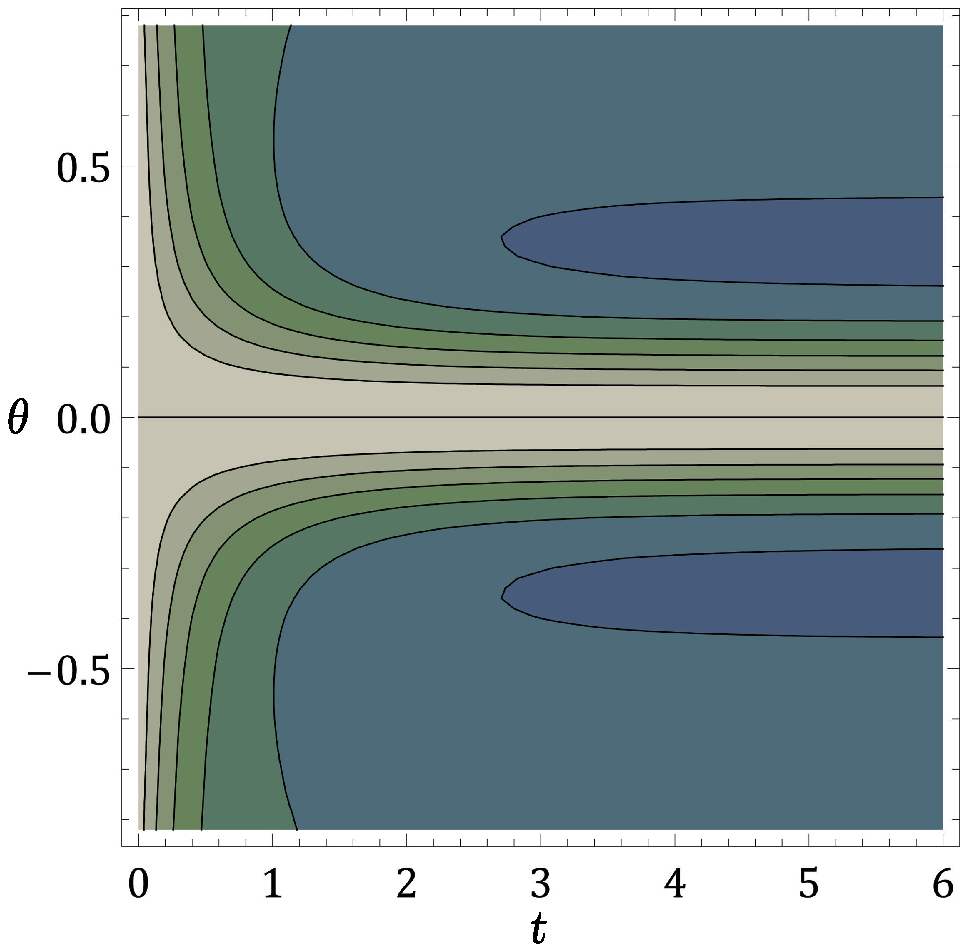}}
 \subfigure[]
   {\includegraphics[width=6.5cm]{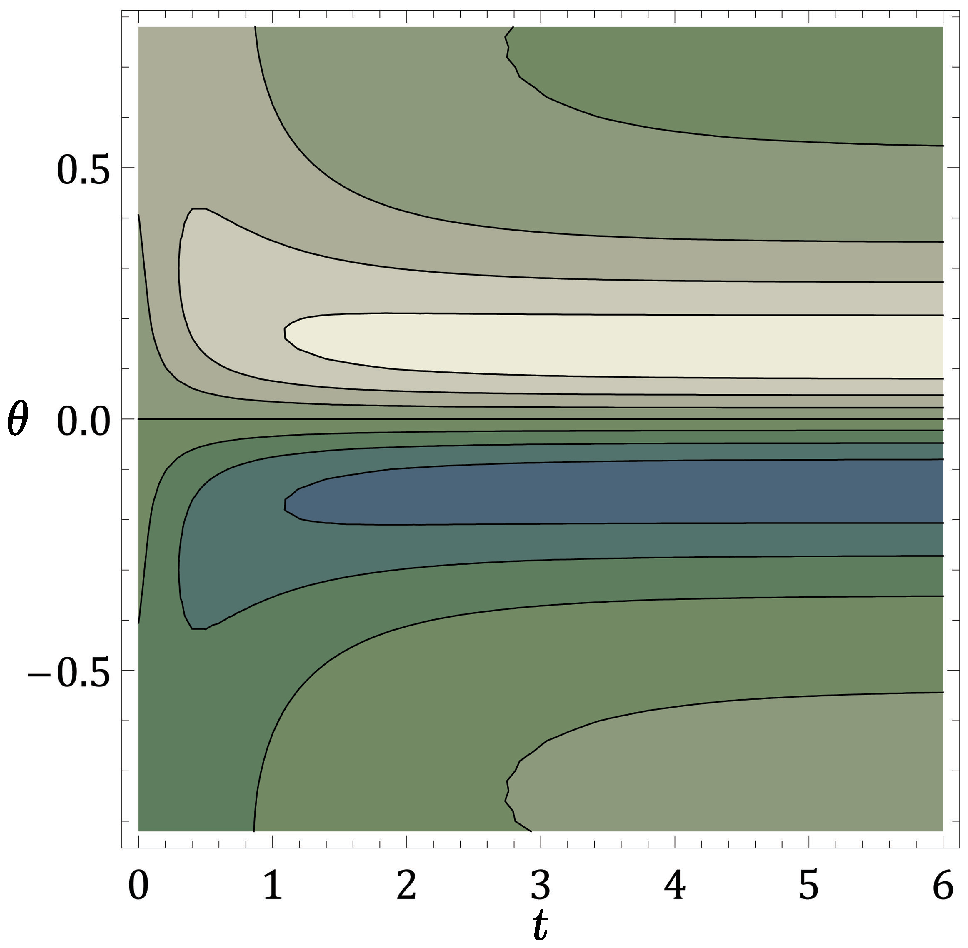}}
 \caption{$\alpha$-stable L\'evy white noise excitation: a) Real part of the CF; b) Imaginary part of the CF; c) Contour plot of $Re(\phi(\theta,t))$; d) Contour plot of $Im(\phi(\theta,t))$ } 
 \label{Figure5}
 \end{figure}

The SFPK equations, that are their spectral counterpart, are instead partial integro-differential equations of the same kind
\begin{subequations}\label{eq34} 
\begin{eqnarray}
\dot{\phi }={\rm{i}}\theta E\left[\left( { - X + \cos \left( {X}\right) }  \right) e^{{\rm{i}}\theta X} \right]-\frac{\theta ^{2} }{2} \phi
\label{eq34a}  
\end{eqnarray}
\begin{eqnarray}
\dot{\phi }={\rm{i}}\theta E\left[\left( { - X + \cos \left( {X}\right) }  \right) e^{{\rm{i}}\theta X} \right]-\lambda {\kern 1pt} {\kern 1pt} \phi \left(1-\phi _{Y} \right)
\label{eq34b}
\end{eqnarray}
\begin{eqnarray}
\dot{\phi }={\rm{i}}\theta E\left[\left( { - X + \cos \left( {X}\right) }  \right) e^{{\rm{i}}\theta X} \right]]-\left|\theta \right|^{\alpha } \phi 
\label{eq34c}
\end{eqnarray}
\end{subequations}
and, by the procedure previously shown, are simply rewritten as 
\begin{equation}
{\boldsymbol {\dot \phi}}  = {i \bf{UT_f}{\boldsymbol {\phi}}} - {\bf{G}{\boldsymbol {\phi}}}
\label{equ35}
\end{equation}
The system of linear differential equations in eq.(\ref{equ35}) under the approximated boundary conditions $\phi({\bar{\theta}},t) = 0$ and proper initial conditions, thus obtaining the approximation of the time evolution of the solutions' characteristic functions for each of the non-linear system. 
Of course, the coefficients defining the matrix $\bf{UT_f}$ are calculated only once because dependent only on the drift. 
These characteristic functionals are complex functions and Figures (\ref{Figure3}), (\ref{Figure4}) and (\ref{Figure5}), show both the real and the imaginary part.

It is worth noting that eqs.(\ref{equ34}) might be solved with other finite difference rules or quadrature formula, but the Lubich's convolution quadrature, core of the method here provided exhibits in general higher numerical stability, accuracy and efficiency \cite{scha06}.

\section{Conclusions}

This paper has shown a numerical procedure to find the statistics of non-linear dynamical systems under L\'evy white noise processes.

First, the Spectral Fokker-Plank equation, which governs the time evolution of the characteristic function, has been rewritten in terms of a Wiener-Hopf integral, indicated as $(T_f\phi)(\theta)$.
This term depends only on the non-linear drift. For every white external excitation, either Gaussian or Poisson or L\'evy $\alpha$-stable, the mathematical structure of the governing equations in terms of the characteristic function remains the same.

This method exploits this fact, providing an efficient scheme to evaluate the Wiener-Hopf integral: to this aim the Lubich's convolution quadrature has been reformulated to deal with this particular case.
The partial integro-differential equations of the characteristic function have thus been transformed into a linear system of ordinary differential equation that can be easily solved. 
In this way the non-stationary solutions of three non-linear stochastic systems have been found and presented to support the method.

Validation and error analysis has also been presented by comparison with a benchmark solution in case of external Gaussian excitation. Moreover, the convergence of the numerical scheme on which the method relies on has also been provided in Appendix B. 

Summing up, the advantages of the proposed method relies on: i) excellent numerical stability given by the Lubich's convolution quadrature method; ii) applicability to every kind of external white noise excitation; iii) ease in constructing the matrices defining the system of linear differential equation that have Toeplitz form.

From the theoretical point of view, it is worth to make some further considerations on the form of the Wiener-Hopf integral $(T_f\phi)(\theta)$ and on its series form found in (\ref{equ26}), in connection to fractional differential operators.

For a long time fractional derivatives and integrals have been considered just particular integral transforms with the property of interpolating ordinary derivatives. 
Properties with respect to Fourier transform, the Gr\"unwald-Letnikov form, the compositions rules, and many others have then supported the use of such a mathematical tool that nowadays is widespread. 

This paper deals with the integral transform $(T_f\phi)(\theta)$ in perfect analogy with the fractional integrals, introducing similar notation and giving the most important properties. This simplifies the use and the mathematical treatment of the specific problem at hand on SDEs. In appendix A the similarities between the most useful properties of fractional calculus and the correspondent properties for the Wiener-Hopf integral transform have also presented.

\section{Acknowledgments}
I wish to deeply thank professor Mario Di Paola of the University of Palermo for his continuous support and encouragement. His inspiring works and contagious passion in research represent an inestimable enduring teaching. 

This work has been carried out while visiting the Engineering Risk Analysis Group of Daniel Straub. I wish to thank Daniel for his support. The grant from Dr. Otto and Karla Likar-Stiftung from the Technical University of Munich is also acknowledged.

\appendix
\section{Considerations on the integral transform $(T_f \phi)(\theta)$} \nonumber

The integral transform $(T_f\phi)(\theta)$ defined in (\ref{eq5a}) is a particular type of convolution integral in which the kernel is the inverse Fourier transform of a given function $f$.
Due to this particular structure of the kernel, many similarities between this integral transform and fractional integrals can be drawn and can be useful in applications. 
In this appendix, by simple application of standard theorems on Fourier transformable functions, these properties are reported.


First, we have shown it this paper that the integral transform
\[
\left( {T_f \phi } \right)\left( \theta  \right) = \int_{ - \infty }^\infty  {\phi \left( u \right)\hat f\left( {u - \theta } \right)} du
\]
by means of a modification of the Lubich's approach can be expressed as 
\[
\left( {T_f \phi } \right)\left( \theta  \right) = \mathop {\lim }\limits_{h \to 0^ +  } \left( {\sum\limits_{k = 0}^\infty  {\alpha _k (f) \phi \left( {\theta  - kh} \right)}  + \sum\limits_{k = 0}^\infty  {\omega _k (f)\phi \left( {\theta  + kh} \right)} } \right)
\]
which reminds the relation between the Riemann-Liouville fractional derivatives and their series approximation by the Gr\"unwald-Letnikov's formula.

$(T_f \phi)(\theta)$ has been defined as integral in the whole real domain, but as like as the right and the left side Riemann-Liouville fractional integrals, it can be also defined in the semi-axis. 
Indeed it might be useful to split $(T_f\phi)(\theta)$ in two operators in which, the first takes into account of all the values preceding $\theta$, indicated by $(T{_f}_+\phi)(\theta)$, while the second takes the values following $\theta$ and is indicated as $(T{_f}_-\phi)(\theta)$. 
More explicitly, the series approximation of these two integrals is plainly obtained by the results of the previous sections and reads
\begin{subequations}\label{equ28}
\begin{eqnarray}
\left( {T{_f}_ +  \phi } \right)\left( \theta  \right) = \lim _{h \to 0} \sum\limits_{k = 0}^\infty  {\alpha _k (f) \phi \left( {\theta  - kh} \right)}
\label{equ28a}
\end{eqnarray}
\begin{eqnarray}
\left( {T{_f}_ -  \phi } \right)\left( \theta  \right) = \lim _{h \to 0} \sum\limits_{k = 0}^\infty  {\omega _k (f)\phi \left( {\theta  + kh} \right)
} 
\label{equ28b}
\end{eqnarray}
\end{subequations}
It is plain that $\left( {T_f \phi } \right)\left( \theta  \right) = \left( {T{_f}_ +  \phi } \right)\left( \theta  \right) + \left( {T{_f}_ -   \phi } \right)\left( \theta  \right)$.

Furthermore, the similitude with the fractional operators suggests how to calculate truncated versions of the operators $\left( {T{_f}_ +  \phi } \right)\left( \theta  \right)$ and $\left( {T{_f}_ -  \phi } \right)\left( \theta  \right)$. 
Let us consider a bounded interval $[\theta_i, \theta_f]$. It is easy to demonstrate that, in perfect analogy with the Gr\"unwald-Letnikov fractional integral defined on this bounded interval, the following relations
\begin{subequations}\label{equ29}
\begin{eqnarray}
\left( {T{_f}_ +  \phi } \right)\left( \theta  \right) = \lim _{h \to 0} \sum\limits_{k = 0}^{\left[ {\frac{{\theta  - \theta_i}}{h}} \right]} {\alpha _k (f)\phi \left( {\theta  - kh} \right)
} 
\label{equ29a}
\end{eqnarray}
\begin{eqnarray}
\left( {T{_f}_ -  \phi } \right)\left( \theta  \right) = \lim _{h \to 0} \sum\limits_{k = 0}^{\left[ {\frac{{\theta_f - \theta }}{h}} \right]} {\omega _k (f)\phi \left( {\theta  + kh} \right)
} 
\label{equ29b}
\end{eqnarray}
\end{subequations}
hold true, where the symbol $[\cdot]$ indicates the integer part. 

Other relevant properties can be derived by plain application of the Fubini's theorem. The notation introduced for the integral $(T_f\phi)(\theta)$ allows to simplify the following relations. 

If we indicate the Fourier transform as ${\cal F}$ and apply it to $(T_f\phi)(\theta)$ we obtain
\begin{subequations}\label{eqF1}
\begin{eqnarray}\label{eqF1a}
{\cal F}\left\{ {\left( {T_f \phi } \right)\left( \theta  \right);\omega } \right\} = {\cal F}\left\{ {\phi \left( \theta  \right);\omega } \right\}f\left( { - \omega } \right)
\end{eqnarray}
\begin{eqnarray}\label{eqF1b}
{\cal F}^{ - 1} \left\{ {\left( {T_f \phi } \right)\left( \theta  \right);\omega } \right\} = {\cal F}^{ - 1} \left\{ {\phi \left( \theta  \right);\omega } \right\}f\left( \omega  \right)
\end{eqnarray}
\end{subequations}

The Fubini's formula allows also to write the equivalent form of the so-called fractional integration by parts formula (\cite{samko}, p.96), that in this case assumes the form
\begin{equation}
\int_{ - \infty }^\infty  {\left( {T_f \phi } \right)\left( \theta  \right)g\left( \theta  \right)d\theta }  = \int_{ - \infty }^\infty  {\left( {T_{f* } g} \right)\left( \theta  \right)\phi \left( \theta  \right)d\theta } 
\label{eqF2}
\end{equation}
where $f* =f(-x)$.

Lastly, analogous form of the composition rule for fractional operators is reported. It can be shown that, by applying the integration by parts formula, above reported, and the Fubini's theorem, the following formula
\begin{equation}
\left( {T_g \left( {T_f \phi } \right)} \right)\left( \theta  \right) = \left( {T_{fg} \phi } \right)\left( \theta  \right) = \int_{ - \infty }^\infty  {\phi \left( u \right)\mathop {fg}\limits^{\_\_\_} \left( {u - \theta } \right)du} 
\label{eqF3}
\end{equation}
holds true, where
\[
\mathop {fg}\limits^{\_\_\_} \left( \theta  \right) = \frac{1}{{2\pi }}\int_{ - \infty }^\infty  {g\left( t \right)f\left( t \right)e^{ - it\theta } dt} 
\]
is the inverse Fourier transform of the product of the functions $f(t)$ and $g(t)$.

\section{On the convergence of (\ref{equ26})} \nonumber
Let us assume that $f(x)$ is Lipschitz and satisfies some growth condition such that the SDEs in (\ref{eq_1}) have an unique solution. 
Moreover, it is assumed that $f(x)$ is Fourier transformable, at least in generalized sense, and that the inverse Laplace transforms (\ref{eq12}) are defined for every $s>s_0$.

The only approximation introduced in section 3 is the Eulerian scheme in equation (\ref{equ18}) that in its general form can be written as
\[
a_0 y_n  + a_1 y_{n - 1}  = h\,b_0 \,\phi \left( {y_n } \right)
\]
and it is known to converge (see, for example \cite{youn72}, p.494) if $\phi$ is Lipschitz, i.e
\[
\left| {\phi \left( \theta  \right) - \phi \left( u \right)} \right| \le L\left| {\theta  - u} \right|
\]
and 
\[
\frac{{hL\left| {b_0 } \right|}}{{\left| {a_0 } \right|}} < 1
\]

It suffices to recall that in our case $a_0 = 1-sh$, $a_1 = -1$ and $b_0 =1$ and to recall that the function $\phi(y)$ is a characteristic function, i.e. $\left| {\phi \left( y \right)} \right| \le 1$ for every real $y$ to derive that the sum in (\ref{equ22}) converges for
\[
s_0  < s < \frac{{1 - h}}{h}
\]

From this relation a proper value of the increment $h$ can be selected for numerical implementation.

\bibliographystyle{plain}
\bibliography{biblio}

\end{document}